\documentclass[prb,draft,preprint,amsfonts,amssymb,amsmath,showpacs]{revtex4}
\begin{document}

\title{Relativistic Model of two-band Superconductivity in (2+1)-dimension}
\author{Tadafumi Ohsaku}
\affiliation{Department of Physics, and Department of Chemistry, Graduate School of Science, 
Osaka University, Toyonaka, Osaka, Japan, 
and Research Center for Nuclear Physics, Osaka University, Ibaraki, Osaka, Japan,
and Yukawa Institute for Theoretical Physics, Kyoto University, Kyoto, Japan.}

\date{\today}

\newcommand{\bmx}{\mbox{\boldmath $x$}}
\newcommand{\bmy}{\mbox{\boldmath $y$}}
\newcommand{\bmk}{\mbox{\boldmath $k$}}
\newcommand{\bmp}{\mbox{\boldmath $p$}}
\newcommand{\bmq}{\mbox{\boldmath $q$}}
\newcommand{\bmP}{\mbox{\boldmath $P$}}  
\newcommand{\kfey}{\ooalign{\hfil/\hfil\crcr$k$}}
\newcommand{\pfey}{\ooalign{\hfil/\hfil\crcr$p$}}
\newcommand{\qfey}{\ooalign{\hfil/\hfil\crcr$q$}}
\newcommand{\Deltafey}{\ooalign{\hfil/\hfil\crcr$\Delta$}}

\begin{abstract}

We investigate the relativistic model of superconductivity in (2+1)-dimension. 
We employ the massless Gross-Neveu model at finite temperature and density, 
to study the superconductivity and superconducting instability. 
Our investigation is related to the superconductivity in (2+1)-dimensional two-band systems
like ${\rm MgB_{2}}$ or intercalated graphite.

\end{abstract}


\maketitle

\section{Introduction}

Condensed matter physics in low-dimension is 
an interesting research area both experimentally and theoretically~[1]. 
For example, the discovery of the high-$T_{c}$ superconductors made 
a large impact on condensed matter physics, and in such systems, 
the superconductivity is considered to occur in a quasi-two-space-dimensional 
copper oxide ${\rm CuO_{2}}$ plane~[2-7]. 
The recent discovery of the superconductivity in ${\rm MgB_{2}}$ 
was also a quite important event~[8], 
and people expect that a plane constructed by ${\rm B}$ atoms plays 
the main role in the superconductivity. Graphite and carbon nanotube also 
attract our attention, and they are plane systems~[9]. 
Therefore, the superconductivity in two-dimensional 
systems is of prime importance, and well recognized in condensed matter physics. 
In the theoretical side, Aoki et al. studied the effect of the dimensionality~[10], 
the effect of band structures~[11], the multiband effect ( two-band model~[12], four-band model~[13] ) 
and the effect of shapes of the Fermi surfaces~[14], 
in superconductivity of several systems 
( see also Ref. 15 for two-band model and Ref. 16 for three-band model ). 
The keywords of recent theoretical investigation into superconductivity 
are "two-space-dimensional" and "two-band."

The two-band models of superconductivity~[17-28] have the origin in 
the papers of Suhl, Matthias and Walker~[17], and Kondo~[18,19], 
and they were applied to several systems under various situations. 
Both ${\rm MgB_{2}}$ and graphite have honeycomb lattice structure, 
and essentially they are two-band systems. 
Quite recently, a theoretical study of two-band superconductivity 
in ${\rm MgB_{2}}$ appeared~[22-24].
The experimental evidence for two-band superconductivity 
in ${\rm MgB_{2}}$ was also obtained~[29]. 
Some relations between superconductivity and 
excitonic state ( exciton condensed state ) in two-band models were also examined~[25,26]. 
Low-energy effective theories for two-band superconductivity similar to 
the Ginzburg-Landau model were proposed~[27,28]. 
Some theoreticians consider that the copper oxide high-$T_{c}$ superconductor 
can also be described by a two-band theory ( the d-p model )~[3,4]. 
Today, much attentions are paid for two-band superconductivity. 
There are various two-band models for superconductivity. 
For example, Suhl et al.  used the model which has only an attractive interaction 
between particles in a two-band system~[17]. 
Yamaji discussed a pairing problem by an interband polarization function 
arised from a repulsive interaction~[20]. 
Kondo found that a kind of two-band effect enhances the superconductivity~[18]. 
However, until now, it is not clear how much is the contribution of the lower band. 
There is no quantitative understanding about the strength of the two-band effect, 
especially the lower-band effect.

Relativistic fermion often appears in condensed matter systems~[30-36]. 
Semenoff studied the (2+1)-dimensional ( 2 for space and 1 for time ) 
relativistic fermion model in graphite. 
Based on the character of (2+1)-dimensional relativistic quantum field theory~[37-41], 
he discussed the anomaly ( the Chern-Simons term and the fractional fermion number )~[30]. 
As a consequence of the honeycomb lattice structure of graphite, 
its band structure has two degeneracy points 
in the first Brillouin zone ( two conical intersections between the upper band and lower band ). 
Then the relativistic fermion model is obtained in the linear dispersion approximation~[30-34]:
\begin{eqnarray}
\epsilon &=& v_{F}\gamma^{0}(\gamma^{1}p_{x}+\gamma^{2}p_{y})
\end{eqnarray}
( as illustrated in Figs. 1 and 2 ). 
Here, $v_{F}$ is the Fermi velocity, and $p_{x}$ and $p_{y}$ are momentum operators. 
By using the Dirac fermion model with a Coulomb repulsion, Gonz\'alez et al. concluded that 
the superconductivity can emerge in graphite~[32]. 
Shankar derived a Dirac fermion model in (2+1)-dimensional doped antiferromagnets~[35]. 
Because those relativistic models have two bands ( positive and negative energy states ), 
they can provide powerful techniques to study the low-energy 
and long-wavelength properties of (2+1)-dimensional two-band systems.

With these considerations given above, 
we investigate the relativistic model of two-band superconductivity in (2+1)-dimension. 
The purposes of this paper is to examine a two-band effect in superconductivity in 
the (2+1)-dimensional system. We concentrate on the examination of the effect of the lower band, 
by extracting the strength of its contribution in the superconductivity. 
In real substances, our theory can be applied to systems which have two-space-dimensional 
honeycomb lattice structure like ${\rm MgB_{2}}$, electron-doped-graphite 
and graphite intercalation compounds ( ${\rm LiC_{6}}$, ${\rm KC_{8}}$, etc. ), 
or to systems of the kagome lattice structure~[42]. 
Because real substances are not exactly (2+1)-dimensional, 
we do not consider the Kosterlitz-Thouless transition in two-space-dimensional 
superconductivity seriously~[43].

This paper is organized as follows. 
In Sec. II, we introduce the Gross-Neveu model as 
our model Lagrangian, and discuss its characteristic aspects. 
By using this model Lagrangian, we study the superconductivity, namely, 
one of the phenomena of dynamical $U(1)$-gauge-symmetry breaking in (2+1)-dimension. 
In Sec. III, the Gor'kov formalism~[44] for a contact attractive interaction 
in our theory is presented. In Sec. IV, the group-theoretical consideration of 
the mean fields ( gap functions ) is provided. 
In Sec. V, by using the Gor'kov formalism, the gap equations in our theory are derived, 
and they are solved numerically. In 1965, Kohn and Luttinger proved that, 
whether a two-body interaction is attractive or repulsive, 
there is a Cooper instability in an interacting many-fermion system ( the Kohn-Luttinger effect )~[45]. 
Shankar and Chubukov independently proved the existence of the Kohn-Luttinger ( KL ) effect 
in a two-space-dimensional system~[46,47]. 
In Sec. VI, for the pairing problem in the case of a repulsive interaction, 
the KL effect in our theory is examined by using the Bethe-Salpeter ( BS ) formalism. 
Finally in Sec VII, we give the conclusion of this work, with further possible investigation in our theory.

\section{The Model Lagrangian and Hamiltonian}

In this section, we discuss the character of the model Lagrangian 
to study the (2+1)-dimensional relativistic superconductivity. 
We take the following Lagrangian for the starting point:
\begin{eqnarray}
{\cal L}(x) &=& \bar{\psi}(x)i\gamma^{\mu}\partial_{\mu}\psi(x)+\frac{G}{2}(\bar{\psi}(x)\psi(x))^{2}.
\end{eqnarray}
This is the (2+1)-dimensional Gross-Neveu model~[48-52]. 
The first term is the kinetic term of the Dirac field, 
$\psi$ and $\bar{\psi}$ are the two-component relativistic spinors 
describing the Dirac fields. Here, we do not give a mass to the fermion. 
It is well known in (2+1)-dimensional relativistic field theory~[37-41], 
the Dirac mass term $m\bar{\psi}\psi$ violates both the parity and the time reversal symmetries. 
As discussed in the introduction, we treat the system which has a degeneracy point at zero momentum. 
We have to consider the massless case. 
In Eq. (2), we introduce the four-body contact interaction at the same spacetime point. 
Here, we consider one of the simplest relativistic interaction which may generate a superconductivity. 
When $G>0$, it will give an attractive interaction. 
We set aside the question of the origin of the attractive interaction, 
and regard the Lagrangian (2) as a phenomenological one. 
On the other hand, when $G<0$, it will become a similar interaction to 
the on-site repulsion of the Hubbard model in its continuum limit. 
If we employ naive power counting analysis, we find that 
the theory is unrenormalizable and we have to introduce a cutoff. 
Until now, the (2+1)-dimensional Gross-Neveu model is studied extensively, 
especially in the context of the dynamical chiral symmetry breaking~[48-52]. 
Rosenstein et al. treated the model by using the 1/N expansion ( here, N is a number of flavor )~[49]. 
The method of 1/N expansion, a kind of mean-field theory, 
treats the sum of an infinite number of Feynman diagrams similar to the Hartree-Fock theory. 
Rosenstein et al. also gave a proof of renormalizability of the model under the 1/N expansion. 
On the other hand, we study a subset of diagrams different from that of Rosenstein et al. 
In fact, Rosenstein et al. investigated the chiral symmetry breaking arised from 
fermion-antifermion condensate $\langle \bar{\psi}\psi\rangle \ne 0$, 
while we treat a fermion-fermion pairing problem, $\langle\psi\psi\rangle \ne 0$. 
In other words, Rosenstein et al. studied the dynamical generation of Dirac mass, 
while we study the dynamical generation of Majorana-type mass. 
In this paper, we will not study the possibility of renormalizability of our theory, 
and treat ultraviolet divergences with a simple cutoff scheme. 
The cutoff indicates the momentum range appropriate to the linear dispersion approximation.
Our Lagrangian should be regarded as a model in this range of momentum space.
For the purpose of this paper, we do not have to introduce the constant for 
the Fermi velocity $v_{F}$ in the model. 
Equation (2) itself has symmetries of Poincar\'{e} invariance, 
$U(1)$-gauge invariance, charge conjugation invariance, 
spatial inversion and time reversal invariance.

For the Clifford algebra~[37,38] of $\gamma$-matrices, using the Pauli matrices:
\begin{eqnarray}
\gamma^{0}=\sigma^{3},\quad \gamma^{1}=i\sigma^{1},\quad \gamma^{2}=i\sigma^{2},
\end{eqnarray}
( here the chirality $\eta\equiv\frac{i}{2}{\rm tr}\gamma^{0}\gamma^{1}\gamma^{2}=+1$ ) we obtain the next relations:
\begin{eqnarray}
\{\gamma^{\mu},\gamma^{\nu}\}=2g^{\mu\nu}, \qquad \gamma^{\mu\dagger}=\gamma^{0}\gamma^{\mu}\gamma^{0}, \qquad \gamma^{\mu}\gamma^{\nu}=g^{\mu\nu}-i\epsilon^{\mu\nu\lambda}g_{\lambda\rho}\gamma^{\rho}.
\end{eqnarray}
Here, $\epsilon^{012}=1$. We take the metric convention as $g^{\mu\nu}={\rm diag}(1, -1, -1)$. The charge conjugation matrix is given as
\begin{eqnarray}
C^{-1}\gamma^{\mu}C=-\gamma^{\mu T}, \qquad C^{\dagger}C =1. 
\end{eqnarray}
For the Poincar\'{e} algebra, the generator of the Lorentz transformation satisfies the $SO(2,1)$ algebra:
\begin{eqnarray}
[j^{\mu},j^{\nu}] &=& -i\epsilon^{\mu\nu\lambda}g_{\lambda\rho}j^{\rho},
\end{eqnarray}
where $j^{0}$ is the generator of two-dimensional rotation as the $U(1)$ phase transformation, 
while $j^{1}$ and $j^{2}$ are the boost operators. 
Especially the representation on the Dirac field is given as
\begin{eqnarray}
j^{\mu}=\frac{1}{2}\gamma^{\mu},
\end{eqnarray}
then $\psi$ transforms as
\begin{eqnarray}
\psi(x) \to e^{i\omega\cdot j}\psi(x) = e^{\frac{i}{2}\omega\cdot\gamma}\psi(x).
\end{eqnarray}

The Hamiltonian of our problem becomes 
\begin{eqnarray}
H &=& \int d^{2}\bmx\bar{\psi}(-i\vec{\gamma}\cdot\vec{\nabla}-\gamma^{0}\mu)\psi -\frac{G}{2}\int d^{2}\mbox{\boldmath $x$}(\bar{\psi}\psi)^{2}. 
\end{eqnarray}
Here, we introduce the chemical potential $\mu$. 
In the relativistic field theory, 
it descibes the finite density at $\mu\ne0$ as the conjugate of 
the particle number minus antiparticle number. 
When we consider a crystal by the model, $\mu$ becomes the conjugate of 
the electron number in upper band minus hole number in lower band. 
$\mu$ determines the position of the Fermi level measured from the degeneracy point.
The system has a Fermi circle at $\mu\ne 0$.
Throughout this study, we completely neglect the temperature dependence of $\mu$, 
and we treat $\mu$ as a parameter introduced from the outside of the system. 
Thus we set $\mu =\epsilon_{F}$ ( Fermi energy ).
The Fermi level of graphite locates on the degeneracy point.
Hence, in the case of intercalated graphite, $\mu$ determines the electron doping concentration.

Before closing this section, we would like to comment on the reason why we use
the theory of the two-component Dirac field. In the theory of (2+1)-dimension,
we have a choice between a two-component Dirac field and a four-component Dirac field.
In fact, the dispersion of the honeycomb lattice has two degeneracy points.
Gonzalez used a four-component theory to take into account the effect of the
interaction between the two Fermi points in graphite~[32]. 
However, the purpose of this paper is to extract the two-band effect,
especially the lower-band effect. For this purpose, we have to construct the model
as simple as possible. The criteria of our model are given as follows:
(1) It will give a conical ( relativistic ) dispersion, (2) it has an interaction
which may arise a superconductivity. 
From these criteria, we consider the Lagrangian (2) as the simplest model. 
If we use a four-component theory of superconductivity, 
it gives a problem of the effect of the interaction between two distinct Fermi circle.
We regard that the effect is essentially different from the lower band effect 
which is studied in the context of this paper.
Hence we use a two-component theory, and treat the problem of a single conical-dispersion system. 
The operator $\gamma^{\mu}p_{\mu}$ has two linearly independent solutions; 
one for positive energy state and another for negative energy state.
Therefore, though $\psi$ is a two-component spinor, there is no degeneracy of a spin-degree of freedom.
Both the upper and lower bands have no spin degeneracy.
We treat a kind of spinless model, and it is enough for our purpose of this paper.

\section{Gor'kov Formalism}

In this section, we derive the Gor'kov theory~[44] for pairing problem 
under the attractive interaction $G>0$ in the Lagrangian (2). 
The formalism given in this section is parallel with the (3+1)-dimensional theory~[53-55]. 
The field equations are obtained from the Lagrangian (2) by the action principle: 
\begin{eqnarray}
0 &=& \frac{\partial{\cal L}}{\partial \bar{\psi}}-\partial_{\mu}\frac{\partial{\cal L}}{\partial(\partial_{\mu} \bar{\psi})} = i\gamma^{\mu}\partial_{\mu}\psi+G(\bar{\psi}\psi)\psi, \\
0 &=& \frac{\partial {\cal L}}{\partial \psi}-\partial_{\mu}\frac{\partial {\cal L}}{\partial (\partial_{\mu}\psi)} = -i\partial_{\mu}\bar{\psi}\gamma^{\mu}+G(\bar{\psi}\psi)\bar{\psi}.
\end{eqnarray}
For the derivation of the Gor'kov formalism, we introduce various propagators. 
We use the 4-component Nambu notation~[56]: 
\begin{eqnarray}
 \hat{\Psi}(x) \equiv \left(
 \begin{array}{c}
 \hat{\psi}(x) \\
\hat{\bar{\psi}}^{T}(x)
\end{array}
\right), \quad \hat{\bar{\Psi}}(x) &\equiv& (\hat{\bar{\psi}}(x),\hat{\psi}^{T}(x)),   
\end{eqnarray}
where $T$ means the transposition. The definition of the one-particle propagator is
\[
{\bf G}(x,y) \equiv -i\langle T\hat{\Psi}(x)\hat{\bar{\Psi}}(y) \rangle
\]
\begin{equation} 
= \left(
 \begin{array}{cc}
 -i\langle T\hat{\psi}_{\alpha}(x)\hat{\bar{\psi}}_{\beta}(y)\rangle & -i\langle T\hat{\psi}_{\alpha}(x)\hat{\psi}^{T}_{\beta}(y)\rangle \\
 -i\langle T\hat{\bar{\psi}}^{T}_{\alpha}(x)\hat{\bar{\psi}}_{\beta}(y)\rangle & -i\langle T\hat{\bar{\psi}}^{T}_{\alpha}(x)\hat{\psi}^{T}_{\beta}(y)\rangle 
\end{array}
\right) 
= \left(
 \begin{array}{cc}
 S_{F}(x,y)_{\alpha\beta} & -iF(x,y)_{\alpha\beta}  \\
 -i\bar{F}(x,y)_{\alpha\beta} & -S_{F}(y,x)_{\beta\alpha}
\end{array}
\right). 
\end{equation}
This is a 4$\times$4 matrix. $T$ means the time-ordered product, 
and $\langle \cdots\rangle$ means the expectation value. 
$S_{F}$ is the Feynman propagator for quasiparticle, 
while $-iF$ and $-i\bar{F}$ are the anomalous propagators. 
Next, we obtain the equations of motion for the propagators (13). 
We employ the Gor'kov factorization in (10) and (11), taking account of only 
the superconducting pair-correlation by introducing the mean-field approximation. 
Then we obtain the relativistically generalized (2+1)-dimensional Gor'kov equation 
written down as a 4$\times$4 matrix equation:
\[
\left(
\begin{array}{cccc}
i\gamma^{\mu}\partial_{\mu}+\gamma^{0}\mu & \Delta(x) \\
\bar{\Delta}(x) & i(\gamma^{\mu})^{T}\partial_{\mu}-(\gamma^{0})^{T}\mu
\end{array}
\right)
\left(
\begin{array}{cccc}
S_{F}(x,y) & -iF(x,y) \\
-i\bar{F}(x,y) & -S_{F}(y,x)^{T}
\end{array}
\right)
\]
\begin{equation} 
=
\left(
\begin{array}{cccc}
\delta^{(3)}(x-y) & 0 \\
0 & \delta^{(3)}(x-y)
\end{array}
\right).
\end{equation} 
$\Delta(x)$ and $\bar{\Delta}(x)$ are 2$\times$2 matrix mean fields, so called order parameters. The definitions are
\begin{eqnarray}
\Delta(x_{0},\mbox{\boldmath $x$})_{\alpha\beta} &\equiv& G F(x^{+}_{0},\mbox{\boldmath $x$};x_{0},\mbox{\boldmath $x$})_{\alpha\beta}  = G\langle\hat{\psi}_{\alpha}(x^{+}_{0},\bmx)\hat{\psi}^{T}_{\beta}(x_{0},\bmx)\rangle,   \\
\bar{\Delta}(x_{0},\bmx)_{\alpha\beta} &\equiv& G \bar{F}(x^{+}_{0},\bmx;x_{0},\bmx)_{\alpha\beta} = G\langle \hat{\bar{\psi}}^{T}_{\alpha}(x^{+}_{0},\bmx)\hat{\bar{\psi}}_{\beta}(x_{0},\bmx)\rangle.
\end{eqnarray}
This gives the self-consistency condition. 
In general, the mean field clearly violates the Lorentz symmetry, as well as the gauge symmetry. 
In other words, the mean field involves quantities other than the scalar.

We will also obtain the Fourier transform of the Gor'kov equation:
\begin{equation}
\left(
\begin{array}{cccc}
\tilde{\kfey} & \Delta \\
\bar{\Delta} & \check{\kfey}^{T}
\end{array}
\right)
\left(
\begin{array}{cccc}
S_{F}(k) & -iF(k) \\
-i\bar{F}(k) & -S_{F}(-k)^{T}
\end{array}
\right)
=
\left(
\begin{array}{cccc}
1 & 0 \\
0 & 1 
\end{array}
\right).
\end{equation}
Here $\tilde{k}\equiv(k_{0}+\mu,\bmk)$ and $\check{k}\equiv(k_{0}-\mu,\bmk)$. $\kfey^{T}$ means the transpose of $\kfey$. The self-consistency condition now becomes
\begin{eqnarray}
\Delta = G\int\frac{d^{3}p}{(2\pi)^{3}}F(p), \qquad \bar{\Delta} = G\int\frac{d^{3}p}{(2\pi)^{3}}\bar{F}(p).
\end{eqnarray}
Here the mean field has only the internal degrees of freedom.

The finite-temperature theory of the Matsubara formalism can be obtained in the same way. We introduce imaginary time $\tau=it$. The temperature Green's function is defined as
\begin{eqnarray}
{\bf {\cal G}}(x,y) &\equiv& -\langle T_{\tau}\hat{\Psi}(x)\hat{\bar{\Psi}}(y)\rangle \nonumber
\end{eqnarray}
\begin{eqnarray}
= 
\left(
\begin{array}{cccc}
-\langle T_{\tau}\hat{\psi}_{\alpha}(x)\hat{\bar{\psi}}_{\beta}(y)\rangle & -\langle T_{\tau}\hat{\psi}_{\alpha}(x)\hat{\psi}^{T}_{\beta}(y)\rangle \\
-\langle T_{\tau}\hat{\bar{\psi}}^{T}_{\alpha}(x)\hat{\bar{\psi}}(y)_{\beta} \rangle & -\langle T_{\tau}\hat{\bar{\psi}}^{T}_{\alpha}(x)\hat{\psi}^{T}_{\beta}(y)\rangle 
\end{array}
\right)
=
\left(
\begin{array}{cccc}
{\cal S}(x,y)_{\alpha\beta} & -{\cal F}(x,y)_{\alpha\beta} \\
-\bar{{\cal F}}(x,y)_{\alpha\beta} & -{\cal S}(y,x)_{\beta\alpha}
\end{array}
\right),
\end{eqnarray}
where $\langle \cdots\rangle$ means the statistical average. From the equation of motion of the temperature Green's function, the Gor'kov equation becomes
\[
\left(
\begin{array}{cccc}
-\gamma^{0}(\frac{\partial}{\partial\tau}-\mu)+i\gamma^{k}\partial_{k} & \Delta(x) \\
\bar{\Delta}(x) & -(\gamma^{0})^{T}(\frac{\partial}{\partial\tau}+\mu)+i(\gamma^{k})^{T}\partial_{k}
\end{array}
\right)
\left(
\begin{array}{cccc}
{\cal S}(x,y)_{\alpha\beta} & -{\cal F}(x,y)_{\alpha\beta} \\
-{\bar {\cal F}}(x,y)_{\alpha\beta} & -{\cal S}(y,x)_{\beta\alpha}
\end{array}
\right)
\]
\begin{equation}
=
\left(
\begin{array}{cccc}
\delta^{(3)}(x-y) & 0 \\
0 & \delta^{(3)}(x-y) 
\end{array}
\right).
\end{equation}
Here the definition of the mean fields are the simple extension of those for the zero temperature:
\begin{eqnarray}
\Delta(\tau,\bmx)_{\alpha\beta} &\equiv& G{\cal F}(\tau^{+},\bmx;\tau,\bmx) = G\langle\hat{\psi}_{\alpha}(\tau^{+},\bmx)\hat{\psi}^{T}_{\beta}(\tau,\bmx)\rangle, \\
\bar{\Delta}(\tau,\bmx)_{\alpha\beta} &\equiv& G{\bar {\cal F}}(\tau^{+},\bmx;\tau,\bmx) = G\langle \hat{\bar{\psi}}^{T}_{\alpha}(\tau^{+},\bmx)\bar{\psi}_{\beta}(\tau,\bmx) \rangle. 
\end{eqnarray}
Fourier transform is also obtained as follows:
\[
\left(
\begin{array}{cccc}
\gamma^{0}(i\omega_{n}+\mu)-\vec{\gamma}\cdot\bmk & \Delta \\
\bar{\Delta} & (\gamma^{0})^{T}(i\omega_{n}-\mu)-(\vec{\gamma})^{T}\cdot\bmk
\end{array}
\right)
\left(
\begin{array}{cccc}
{\cal S}(\omega_{n},\bmk) & -{\cal F}(\omega_{n},\bmk) \\
-{\bar{\cal F}}(\omega_{n},\bmk) & -{\cal S}(-\omega_{n},-\bmk)^{T}
\end{array}
\right)
\]
\begin{equation}
=
\left(
\begin{array}{cccc}
1 & 0 \\
0 & 1
\end{array}
\right).
\end{equation}
Here $\beta\equiv(k_{B}T)^{-1}$ ( $k_{B}$; the Boltzmann constant ), $\omega_{n}=(2n+1)\pi/\beta$ is a fermion discrete frequency. 
Solving Eq. (23), we will obtain the solutions of (17) in the same form, 
except that we need to substitute $k_{0}\to i\omega_{n}$.

\section{Group-theoretical Consideration of the Mean-fields}

Now we perform the group-theoretical consideration of the mean fields. Under the Lorentz transformation:
\begin{eqnarray}
\psi'(x') &=& S\psi(x) = \exp(\frac{i}{2}\omega_{\nu}\gamma^{\nu})\psi(x), 
\end{eqnarray}
then the mean field is transformed as
\begin{eqnarray}
\Delta'(x') &=& \langle\psi'(x')\psi^{'T}(x')\rangle = \langle S\psi(x)\psi^{T}(x)S^{T}\rangle \nonumber \\
&=& S\Delta(x)S^{T} \nonumber \\
&\cong& (1+\frac{i}{2}\omega_{\nu}\gamma^{\nu})\Delta(x)(1+\frac{i}{2}\omega_{\nu}\gamma^{\nu T}) \nonumber \\
&=& \Delta(x)+\frac{i}{2}\omega_{\nu}[\gamma^{\nu},\Delta(x)C^{-1}]C. 
\end{eqnarray}
Thus we can decompose the mean fields as follows:
\begin{eqnarray}
\Delta = (\Delta^{S}+\Delta^{V}_{\mu}\gamma^{\mu})C, \qquad \bar{\Delta} = -C^{-1}(\Delta^{S*}+\Delta^{V*}_{\mu}\gamma^{\mu}),
\end{eqnarray}
where $S$ indicates scalar, while $V$ indicates vector. Under the discrete transformations:
\begin{eqnarray}
\psi \stackrel{\cal C}{\to} C\bar{\psi}^{T} = i\gamma^{1}\gamma^{0}\bar{\psi}^{T} \quad &;& \quad \bar{\psi} \stackrel{\cal C}{\to} -\psi^{T}C^{-1} = \psi^{T}i\gamma^{1}\gamma^{0}, \\ 
\psi(x_{0},\bmx) \stackrel{\cal P}{\to} -i\gamma^{1}\psi(x_{0},\bmx')\quad &;& \quad \bar{\psi}(x_{0},\bmx) \stackrel{\cal P}{\to} -\bar{\psi}(x_{0},\bmx')(-i\gamma^{1}), \\
\psi(x_{0}) \stackrel{\cal T}{\to} -i\gamma^{2}\psi(-x_{0}) \quad &;& \quad \bar{\psi}(x_{0}) \stackrel{\cal T}{\to} -\bar{\psi}(-x_{0})(-i\gamma^{2}),
\end{eqnarray}
( here $\bmx=(x_{1}, x_{2}$) and $\bmx'=(-x_{1}, x_{2})$ ), where ${\cal C}$, ${\cal P}$ and ${\cal T}$ denote the operations of charge conjugation, 
spatial inversion and time reversal, respectively. 
Therefore, the mean fields are transformed as
\begin{eqnarray}
\langle\psi\psi^{T}\rangle \stackrel{\cal C}{\to} C\langle\bar{\psi}^{T}\bar{\psi}\rangle C^{-1} = -\gamma^{2}\langle\bar{\psi}^{T}\bar{\psi}\rangle\gamma^{2}, \\
\langle\bar{\psi}^{T}\bar{\psi}\rangle \stackrel{\cal C}{\to} C\langle\psi\psi^{T}\rangle C^{-1} = -\gamma^{2}\langle\psi\psi^{T}\rangle\gamma^{2}, \\
\langle\psi(x_{0},\bmx)\psi^{T}(x_{0},\bmx)\rangle \stackrel{\cal P}{\to} -\gamma^{1}\langle\psi(x_{0},\bmx')\psi^{T}(x_{0},\bmx')\rangle\gamma^{1}, \\
\langle\bar{\psi}^{T}(x_{0},\bmx)\bar{\psi}(x_{0},\bmx)\rangle \stackrel{\cal P}{\to} -\gamma^{1}\langle\bar{\psi}^{T}(x_{0},\bmx')\bar{\psi}(x_{0},\bmx')\rangle\gamma^{1}, \\
\langle\psi\psi^{T}\rangle \stackrel{\cal T}{\to} \gamma^{2}\langle\psi\psi^{T}\rangle^{*}\gamma^{2}, \\
\langle\bar{\psi}^{T}\bar{\psi}\rangle \stackrel{\cal T}{\to} \gamma^{2}\langle\bar{\psi}^{T}\bar{\psi}\rangle^{*}\gamma^{2}.
\end{eqnarray}
Thus each type of the mean fields is transformed under the spatial inversion and time reversal as
\begin{eqnarray}
\Delta^{S}\gamma^{2} &\stackrel{\cal P}{\to}&  -\Delta^{S}\gamma^{2}, \\
&\stackrel{\cal T}{\to}&  -\Delta^{S*}\gamma^{2}, \\
\Delta^{V}_{0}\gamma^{0}\gamma^{2} &\stackrel{\cal P}{\to}&  \Delta^{V}_{0}\gamma^{0}\gamma^{2}, \\
&\stackrel{\cal T}{\to}&  \Delta^{V*}_{0}\gamma^{0}\gamma^{2}, \\
\Delta^{V}_{1}\gamma^{1}\gamma^{2} &\stackrel{\cal P}{\to}&  -\Delta^{V}_{0}\gamma^{1}\gamma^{2}, \\
&\stackrel{\cal T}{\to}&  \Delta^{V*}_{1}\gamma^{1}\gamma^{2}, \\
\Delta^{V}_{2}\gamma^{2}\gamma^{2} &\stackrel{\cal P}{\to}&  \Delta^{V}_{2}\gamma^{2}\gamma^{2}, \\
&\stackrel{\cal T}{\to}&  -\Delta^{V*}_{2}\gamma^{2}\gamma^{2}.
\end{eqnarray}
Therefore, with 2-dimensional rotation and parity, the mean field is decomposed into 
three irreducible representations: $\Delta^{S}$, $\Delta^{V}_{0}$ 
and $(\Delta^{V}_{1},  \Delta^{V}_{2})$. As expected, $\Delta^{S}$ violates 
the parity similar to the Dirac mass term $m\bar{\psi}\psi$.

\section{The Gap Equation}

In this section, we derive gap equations, and solve them numerically. 
For this purpose, first we have to solve the Gor'kov equation. 
Similar to the case of the (3+1)-dimensional theory~[53-55], 
it is difficult to solve Eq. (17) or Eq. (23) completely in analytical form 
because of its matrix structure. 
Therefore we have to solve the equations assuming the type of the mean field that might be realized. 
Then we obtaine three Gor'kov equations for each type of the mean fields. 
These equations can be solved in the same way as the case of the (3+1)-dimensinal theory~[53-55]. 
We give the following results. First the case of the scalar $\Delta^{S}$:
\[
\left(
\begin{array}{cccc}
S_{F}(k) & -iF(k) \\
-i\bar{F}(k) & -S_{F}(-k)^{T}
\end{array}
\right)
\]
\begin{equation}
=\frac{1}{D(k)}
\left(
\begin{array}{cccc}
(\tilde{\kfey}\check{\kfey}-|\Delta^{S}|^{2})\check{\kfey} & \Delta^{S}(\tilde{\kfey}\check{\kfey}-|\Delta^{S}|^{2})C \\
-\Delta^{S*}C^{-1}(\check{\kfey}\tilde{\kfey}-|\Delta^{S}|^{2}) & -C^{-1}(\check{\kfey}\tilde{\kfey}-|\Delta^{S}|^2)\tilde{\kfey}C 
\end{array}
\right),
\end{equation}
\begin{eqnarray}
D(k) &=& (\tilde{k}\cdot\tilde{k})(\check{k}\cdot\check{k})-2|\Delta^{S}|^{2}(\tilde{k}\cdot\check{k})+|\Delta^{S}|^{4}.
\end{eqnarray}
Next the case of 0th-component of vector $\Delta^{V}_{0}$:
\[
\left(
\begin{array}{cccc}
S_{F}(k) & -iF(k) \\
-i\bar{F}(k) & -S_{F}(-k)^{T}
\end{array}
\right)
\]
\begin{equation}
=\frac{1}{D(k)}
\left(
\begin{array}{cccc}
(\tilde{\kfey}\gamma^{0}\check{\kfey}\gamma^{0}-|\Delta^{V}_{0}|^{2})\gamma^{0}\check{\kfey}\gamma^{0} & \Delta^{V}_{0}(\tilde{\kfey}\gamma^{0}\check{\kfey}-|\Delta^{V}_{0}|^{2}\gamma^{0})C \\
-\Delta^{V*}_{0}C^{-1}(\check{\kfey}\gamma^{0}\tilde{\kfey}-|\Delta^{V}_{0}|^{2}\gamma^{0}) & -C^{-1}(\check{\kfey}\gamma^{0}\tilde{\kfey}\gamma^{0}-|\Delta^{V}_{0}|^2)\gamma^{0}\tilde{\kfey}\gamma^{0}C 
\end{array}
\right),
\end{equation}
\begin{eqnarray}
D(k) &=& (\tilde{k}\cdot\tilde{k})(\check{k}\cdot\check{k})-2|\Delta^{V}_{0}|^{2}(\tilde{k}\cdot\check{k}+2\bmk^{2})+|\Delta^{V}_{0}|^{4}.
\end{eqnarray}
The case of 1st-component of vector $\Delta^{V}_{1}$:
\[
\left(
\begin{array}{cccc}
S_{F}(k) & -iF(k) \\
-i\bar{F}(k) & -S_{F}(-k)^{T}
\end{array}
\right)
\]
\begin{equation}
=\frac{1}{D(k)}
\left(
\begin{array}{cccc}
(\tilde{\kfey}\gamma^{1}\check{\kfey}\gamma^{1}-|\Delta^{V}_{1}|^{2})\gamma^{1}\check{\kfey}\gamma^{1} & \Delta^{V}_{1}(\tilde{\kfey}\gamma^{1}\check{\kfey}+|\Delta^{V}_{1}|^{2}\gamma^{1})C \\
-\Delta^{V*}_{1}C^{-1}(\check{\kfey}\gamma^{1}\tilde{\kfey}+|\Delta^{V}_{1}|^{2}\gamma^{1}) & -C^{-1}(\check{\kfey}\gamma^{1}\tilde{\kfey}\gamma^{1}-|\Delta^{V}_{1}|^2)\gamma^{1}\tilde{\kfey}\gamma^{1}C 
\end{array}
\right),
\end{equation}
\begin{eqnarray}
D(k) &=& (\tilde{k}\cdot\tilde{k})(\check{k}\cdot\check{k})-2|\Delta^{V}_{1}|^{2}(\tilde{k}\cdot\check{k}+2k^{2}_{1})+|\Delta^{V}_{1}|^{4}.
\end{eqnarray}
In all cases, $D(k)$ is second order in $k^{2}_{0}$, and we can easily factorized it as $D(k)=(k_{0}-E_{+})(k_{0}+E_{+})(k_{0}-E_{-})(k_{0}+E_{-})$. 
Here, $E_{+}$ corresponds to the energy of the quasiparticles coming from the upper band ( positive energy states ), 
while $E_{-}$ corresponds to the energy of the quasiparticles coming from the lower band ( negative energy states ).  
Because (44), (46) and (48) are $4\times4$ matrices, there is no degeneracy in these branches of the dispersion relations.
This case relates to the fact that we treat a kind of spinless model.

Now we construct the gap equations by using the Green's functions we have obtained. 
We use the finite-temperature Matsubara formalism. 
From the self-consistency conditions:
\begin{eqnarray}
\Delta &=& G\sum_{n}\frac{1}{\beta}\int\frac{d^{2}\bmk}{(2\pi)^{2}}{\cal F}(\omega_{n},\bmk),
\end{eqnarray}
we obtain the gap equation for the case of the scalar $\Delta^{S}$:
\begin{eqnarray}
1 &=& \frac{G}{2}\int\frac{d^{2}\bmk}{(2\pi)^{2}}\Bigl(\frac{1}{2E_{+}}\tanh\frac{\beta}{2}E_{+}+\frac{1}{2E_{-}}\tanh\frac{\beta}{2}E_{-}\Bigr), \\
E_{\pm} &=& \sqrt{(|\bmk|\mp\mu)^{2}+|\Delta^{S}|^{2}}.
\end{eqnarray}
The second term in the integrand is the contribution coming from 
negative energy states and/or the lower band. 
In the context of this paper, the relativistic effect is the two-band effect. 
For the case of the 0th-component of vector $\Delta^{V}_{0}$, we get
\begin{eqnarray}
1 &=& \frac{G}{2}\int\frac{d^{2}\bmk}{(2\pi)^{2}}\frac{1}{2\sqrt{\mu^{2}+|\Delta^{V}_{0}|^{2}}}\Bigl(-\tanh\frac{\beta}{2}E_{+}+\tanh\frac{\beta}{2}E_{-}\Bigr), \\
E_{\pm} &=& |\bmk|\mp\sqrt{\mu^{2}+|\Delta^{V}_{0}|^{2}},
\end{eqnarray}
and the case of the 1st-component of vector $\Delta^{V}_{1}$, we get
\begin{eqnarray}
-1 &=& \frac{G}{2}\int\frac{d^{2}\bmk}{(2\pi)^{2}}\Bigl((\frac{\sqrt{\bmk^{2}\mu^{2}+|\Delta^{V}_{1}|^2k^{2}_{1}}-k^{2}_{1}}{\sqrt{\bmk^{2}\mu^{2}+|\Delta^{V}_{1}|^{2}k^{2}_{1}}})\frac{1}{2E_{+}}\tanh\frac{\beta}{2}E_{+} \nonumber \\
& & \qquad +(\frac{\sqrt{\bmk^{2}\mu^{2}+|\Delta^{V}_{1}|^{2}k^{2}_{1}}+k^{2}_{1}}{\sqrt{\bmk^{2}\mu^{2}+|\Delta^{V}_{1}|^{2}k^{2}_{1}}})\frac{1}{2E_{-}}\tanh\frac{\beta}{2}E_{-}\Bigr), \\
E_{\pm} &=& \sqrt{\bmk^{2}+\mu^{2}+|\Delta^{V}_{1}|^{2}\mp\sqrt{\bmk^{2}\mu^{2}+|\Delta^{V}_{1}|^{2}k^{2}_{1}}}.
\end{eqnarray}
Integrals of Eqs. (51), (53) and (55) give positive quantities. 
Therefore, from these equations given above, 
we find there are possibilities to obtain the nontrivial solutions for 
the cases of the scalar $\Delta^{S}$ and 0th-component of vector $\Delta^{V}_{0}$. 
To examine whether these equations have nontrivial solutions or not, we have to check them more in detail. 
We will treat them numerically to study the characters of the solutions of these equations. 
We cannot obtain nontrivial solutions for the case of the spatial components of vector $\Delta^{V}_{1}$ 
and $\Delta^{V}_{2}$, because the gap equations (55) become the form "$-1=$ positive quantity".

For the integration of our gap equations, we take
\begin{eqnarray}
\int\frac{d^{2}\bmk}{(2\pi)^{2}} \cdots &=& \frac{1}{4\pi^{2}}\int^{\Lambda}_{0}kdk\int^{2\pi}_{0}d\theta\cdots.
\end{eqnarray}
Therefore, our gap equations include four parameters: Coupling constant $G$, 
chemical potential $\mu$, momentum cutoff $\Lambda$ and temperature $T$. 
Because our theory treats a massless case, there is no unit of energy in our theory. 
To see the effect of the lower band, we also treat the gap equation of the "no sea" 
( neglect the contribution of the lower band (Dirac sea) in the gap equation of the scalar ) case:
\begin{eqnarray}
1 &=& \frac{G}{2}\int^{\Lambda}_{0}\frac{d^{2}\bmk}{(2\pi)^{2}}\frac{1}{2E_{+}}\tanh\frac{\beta}{2}E_{+}, \\
E_{+} &=& \sqrt{(|\bmk|-\mu)^{2}+|\Delta^{no-sea}|^{2}}.
\end{eqnarray}

There is a simple scaling relation in our gap equations. 
When we transform the gap equation in next relations:
\begin{eqnarray}
|\bmk|/\mu=|\bmk'|, \quad \Lambda/\mu=\Lambda', \quad \beta\mu=\beta', \quad |\Delta|/\mu=|\Delta'|,
\end{eqnarray} 
Eq.(51) is transformed as 
\begin{eqnarray}
1 &=& \frac{G}{8\pi}\mu\int^{\Lambda'}_{0} k'dk' \Bigl(\frac{1}{\sqrt{(|\bmk'|-1)^{2}+|\Delta'|^{2}}}\tanh\frac{\beta'}{2}\sqrt{(|\bmk'|-1)^{2}+|\Delta'|^{2}}  \nonumber \\ 
&+& \frac{1}{\sqrt{(
|\bmk'|+1)^{2}+|\Delta'|^{2}}}\tanh\frac{\beta'}{2}\sqrt{(|\bmk'|+1)^{2}+|\Delta'|^{2}}
\Bigr).
\end{eqnarray}
Therefore, when the ratio $\Lambda/\mu$ is fixed and 
when we take $G'=G\mu=const.$, we treat the same equation with (61).

Usually, a gap is much smaller than the Fermi energy: $|\Delta|\ll\mu$. 
For example, in the case of solid, the ratio $|\Delta(T=0)|/\mu$ is $10^{-4}-10^{-2}$. 
We have to choose model parameters $G$ and $\Lambda$ to satisfy the condition. 
In principle, we can choose $\Lambda$ up to the energy scale 
where the conical dispersion is a good approximation. 
Both the Fermi momentum $k_{F}$ and the upper bound of $\Lambda$ should be 
in the order of the inverse of the lattice constant $a^{-1}$. 
Here, we choose the momentum cutoff as $\Lambda=2k_{F}=2\mu$,
and fix it throughout this study. 
After choosing the values of parameters, the integration in the gap equation is performed, 
and search the self-consistency condition under the variation with respect to the amplitude of the gap. 
We have performed the integration in our gap equations by using 
the numerical package {\it Mathematica} version 4.1.

Figure 3 shows the gap at $T=0$ as a function of the coupling constant $G$. 
Here we take the momentum cutoff $\Lambda$ as $\Lambda/2\mu=1$ and we set $\mu=1$. 
Similar to the usual nonrelativistic BCS theory, both the scalar and "no sea" depend 
exponentially on $G$. The gap of the scalar is always larger than the "no sea" case. 
The ratio $|\Delta^{S}(T=0)|/|\Delta^{no-sea}(T=0)|$ is almost always 1.57.

In Fig. 4, we show the temperature $T$ dependence of the solutions in 
the scalar and "no sea" cases. Here we set $G=2\mu^{-1}$, $\mu=1$ and $\Lambda/2\mu=1$. 
In the cases of graphite or ${\rm MgB_{2}}$, 
the relativistic model can be applied to the region of the energy width almost $1-2$eV. 
If we take $\mu=1{\rm eV}$ ( $\Lambda=2{\rm eV}$ ), the critical temperature $T_{c}$ becomes 38K (0.0033eV) 
for the scalar and 24K (0.0021eV) for the "no sea". The ratio $T^{scalar}_{c}/T^{no-sea}_{c}$ is 1.57. 
Both the scalar and "no sea" cases fulfill the BCS universal constant $|\Delta(T=0)|/T_{c}=1.76$~[57]. 
Therefore, both cases obey the BCS-like temperature dependence. 

From our numerical results, we conclude that, 
the scalar ( the two-band case ) gives always larger solution than the "no sea" ( the one-band case ). 
The contribution coming from the lower band  ( the negative energy states ) enhances 
the superconducting gap. It should be noted that, 
the values of the amplitude of the gap and the critical temperature do not predict 
the superconductivity in real substances directly, 
because it is impossible to remove the arbitrariness of the choice of the values of 
the model parameters completely. However, when we consider the two-band system, 
we have to keep in mind the fact that the lower band gives a sizable effect in the superconductivity.

About the 0th-component of vector case, we could not find a reasonable solution 
in our numerical calculation. The reason of this result should be understood in the following way. 
We rewrite Eq. (53) at $T_{c}$ as follows:
\begin{eqnarray}
1 &=& \frac{G}{8\pi}\int^{\Lambda}_{0}pdp\frac{1}{\mu}\Bigl(-\tanh\frac{\beta}{2}(p-\mu)+\tanh\frac{\beta}{2}(p+\mu)\Bigr) \nonumber \\
&=& \frac{G}{4\pi\mu}\int^{\Lambda}_{0}pdp\Bigl( \frac{1}{e^{\beta(p-\mu)}+1}-\frac{1}{e^{\beta(p+\mu)}+1} \Bigr).  
\end{eqnarray}
The integral of the right hand side gives the conserved charge of the system.
Therefore, the equation for determination of $T_{c}$ has no temperature dependence:
We cannot determine $T_{c}$ by Eq.(62). In the (3+1)-dimensional massive theory, 
the 0th-component of vector pairing gives a meaningful state, 
and we discussed various aspects of the relation between the scalar, 
"no sea", "nonrelativistic" and 0th-component of vector~[53-55]. 
This fact contrast with the results of the (2+1)-dimensional theory.

\section{The Kohn-Luttinger Effect}

Now we consider the pairing problem with the repulsive interaction $G<0$ in the Lagrangian (2). 
For this problem, we should examine the appearance of a pole in 
the 4-point function ( the 2-particle Green's function, see Fig. 5(a) ) by using 
the Bethe-Salpeter ( BS ) formalism~[44,58], like the work of Kohn and Luttinger~[45]. 
We start with introducing the fermion-fermion BS equation~[59] in 
the finite-temperature Matsubara formalism: 
\begin{eqnarray}
\chi(\bmp) &=& \sum_{n}\frac{1}{\beta}\int\frac{d^{2}\bmp'}{(2\pi)^{2}}\tilde{\Gamma}(\bmp,\bmp'){\cal S}(\omega_{n},\bmp')\chi(\bmp'){\cal S}(-\omega_{n},-\bmp')^{T}.
\end{eqnarray}
We treat (63) as a function of temperature $T$, and the critical temperature $T_{c}$ is 
determined when a self-consistent solution of $\chi(\bmp)$ appear in Eq.(63). 
The normal-state fermion propagators are given as follows:
\begin{eqnarray}
{\cal S}(\omega_{n},\bmp) &=& -\frac{1}{\tilde{\pfey}}, \qquad {\cal S}(-\omega_{n},-\bmp)^{T} = -C^{-1}\frac{1}{\check{\pfey}}C.  
\end{eqnarray}
Here $\tilde{p}=(i\omega_{n}+\mu,\bmp)$ and $\check{p}=(i\omega_{n}-\mu,\bmp)$. $\tilde{\Gamma}(\bmp,\bmp')$ is the irreducible vertex part.  
The BS amplitude $\chi(\bmp)$ is decomposed as follows:
\begin{eqnarray}
\chi(\bmp) &=& (\chi^{S}(\bmp)+\chi^{V}_{\mu}(\bmp)\gamma^{\mu})C \equiv (\sum^{4}_{A=1}\chi_{A}\Gamma^{A})C. 
\end{eqnarray}
Here, $S$ and $V$ denote scalar and vector, respectively.
$\chi(\bmp)$ has to fulfill the Pauli principle:
\begin{eqnarray}
\chi_{\alpha\beta}(\bmp) &=& -\chi_{\beta\alpha}(-\bmp).
\end{eqnarray}
Then, each component should obey the following relations:
\begin{eqnarray}
\chi^{S}(\bmp)C_{\alpha\beta} = -\chi^{S}(-\bmp)C_{\beta\alpha}, &\qquad\qquad& \chi^{V}_{\mu}(\bmp)(\gamma^{\mu}C)_{\alpha\beta} = -\chi^{V}_{\mu}(-\bmp)(\gamma^{\mu}C)_{\beta\alpha}.
\end{eqnarray}
Therefore we get
\begin{eqnarray}
\chi^{S}(-\bmp) = \chi^{S}(\bmp), &\qquad\qquad& \chi^{V}_{\mu}(-\bmp) = -\chi^{V}_{\mu}(\bmp). 
\end{eqnarray}
In Eq. (63), replace $\chi(\bmp)$ to the expanded form ( Eq. (65) ), 
and take trace of both sides, we obtain the following form:
\begin{eqnarray}
\chi_{A}(\bmp)\frac{1}{2}{\rm tr}(\Gamma^{A}\Gamma^{A}) &=& \sum_{n}\frac{1}{\beta}\int\frac{d^{2}\bmp'}{(2\pi)^{2}}\tilde{\Gamma}(\bmp,\bmp')\frac{1}{2}{\rm tr}\Bigl(\Gamma^{A}{\cal S}(\omega_{n},\bmp')\chi_{A}(\bmp')\Gamma^{A}{\cal S}(-\omega_{n},-\bmp')^{T}\Bigr). 
\end{eqnarray}
Similar to the treatment of the Gor'kov equation in Sec.V, 
we completely neglect couplings between different types of pairing functions. 
Then, for specific $A$, we obtain the following equations:
\begin{eqnarray}
\chi^{S}(\bmp) &=& \frac{1}{2}\int\frac{d^{2}\bmp'}{(2\pi)^{2}}\tilde{\Gamma}(\bmp,\bmp')\chi^{S}(\bmp') \nonumber \\ 
& & \times\Bigl(\frac{1}{2(|\bmp'|-\mu)}\tanh\frac{\beta}{2}(|\bmp'|-\mu)+\frac{1}{2(|\bmp'|+\mu)}\tanh\frac{\beta}{2}(|\bmp'|+\mu)\Bigr), \\
\chi^{V}_{0}(\bmp) &=& \frac{1}{2}\int\frac{d^{2}\bmp'}{(2\pi)^{2}}\tilde{\Gamma}(\bmp,\bmp')\chi^{V}_{0}(\bmp')\frac{1}{2\mu}\Bigl(-\tanh\frac{\beta}{2}(|\bmp'|-\mu)+\tanh\frac{\beta}{2}(|\bmp'|+\mu)\Bigr), \\
-\chi^{V}_{1}(\bmp) &=& \frac{1}{2}\int\frac{d^{2}\bmp'}{(2\pi)^{2}}\tilde{\Gamma}(\bmp,\bmp')\chi^{V}_{1}(\bmp')\frac{1}{2\mu|\bmp'|} \nonumber \\
& & \times\Bigl(\frac{(p'_{1})^{2}-\mu|\bmp'|}{|\bmp'|-\mu}\tanh\frac{\beta}{2}(|\bmp'|-\mu)-\frac{(p'_{1})^{2}+\mu|\bmp'|}{|\bmp'|+\mu}\tanh\frac{\beta}{2}(|\bmp'|+\mu)\Bigr). 
\end{eqnarray}
In this paper, we estimate $\tilde{\Gamma}(\bmp,\bmp')$ by using 
a random phase approximation ( RPA ) ( as illustrated in Fig. 5(b) ). 
The polarization is given as
\begin{eqnarray}
\Pi(\omega_{l},\bmq) &=& \sum_{n}\frac{1}{\beta}\int\frac{d^{2}\bmp}{(2\pi)^{2}}{\rm tr}{\cal S}(\omega_{l}+\omega_{n},\bmp+\bmq){\cal S}(\omega_{n},\bmp). 
\end{eqnarray}
Here, $\omega_{l}$ is a boson discrete frequency. After summing up $\omega_{n}$, we negrect $\omega_{l}$ dependence of $\Pi(\omega_{l},\bmq)$ to obtain the static polarization:
\begin{eqnarray}
\Pi(0,\bmq) &=& \int\frac{d^{2}\bmp}{(2\pi)^{2}}\frac{1}{\epsilon^{2}_{\bmp+\bmq}-\epsilon^{2}_{\bmp}} \nonumber \\
& & \times \Bigg\{ \frac{\bmp\cdot\bmq}{\epsilon_{\bmp}}\Bigl(\frac{1}{e^{\beta(\epsilon_{\bmp}-\mu)}+1}+\frac{1}{e^{\beta(\epsilon_{\bmp}+\mu)}+1} -1\Bigr) \nonumber \\
& & +\frac{(\bmp+\bmq)\cdot\bmq}{\epsilon_{\bmp+\bmq}}\Bigl(\frac{1}{e^{\beta(\epsilon_{\bmp+\bmq}-\mu)} +1}+\frac{1}{e^{\beta(\epsilon_{\bmp+\bmq}+\mu)} +1} -1 \Bigr) \Bigg\},
\end{eqnarray}
where, $\epsilon_{\bmp}=|\bmp|$ is the relativistic dispersion for massless particle. 
The integration in (74) has a ultraviolet divergence, and we have to introduce a momentum cutoff $\Lambda$. 
To obtain $\Pi(0,\bmq)$, we perform the integration numerically in Eq. (74). 
Figure 6 shows the results of numerical integration for $\Pi(0,|\bmq|)$ with several $T$. 
Here we set $\Lambda=2p_{F}=2\mu$ with $\mu=1$. We find a peak near $|\bmq|=2p_{F}$. 
This behavior is a reflection of the sharpness of the Fermi surface. 
The peak decreases when $T$ increases. 
To treat the Cooper problem, we concentrate on the behavior of $\Pi(0,|\bmq|)$ at the Fermi surface. 
To see this, we define $\bmq=\bmp'-\bmp$ and substitute $|\bmq|$ to $|\bmp'-\bmp|$ in $\Pi(0,|\bmq|)$. 
Then we set $|\bmp|=|\bmp'|=p_{F}$, 
we get $\Pi(0,p_{F},p_{F},\cos\theta_{\hat{\bmp}\cdot\hat{\bmp}'})$. 
Here $\theta_{\hat{\bmp}\cdot\hat{\bmp}'}=\theta_{\hat{\bmp}}-\theta_{\hat{\bmp}'}$. 
We calculate $\Pi(0,p_{F},p_{F},\cos\theta_{\hat{\bmp}\cdot\hat{\bmp}'})$ numerically. 
Figure 7 shows the angular dependence of $\Pi(0,p_{F},p_{F},\cos\theta_{\hat{\bmp}\cdot\hat{\bmp}'})$ 
with several $T$. The dent at $\theta_{\hat{\bmp}\cdot\hat{\bmp}'}=\pi$ grows when $T$ incleases. 
At enough low temperatures, the angular dependence is  almost $\cos\theta$, 
and then we can use the following expression to examine the pairing properties of the system:
\begin{eqnarray}
\Pi(0,p_{F},p_{F},\cos\theta_{\hat{\bmp}\cdot\hat{\bmp}'})_{T=0} &\simeq& \frac{1}{2\pi}(\Lambda-p_{F})+\alpha(1-\cos\theta_{\hat{\bmp}\cdot\hat{\bmp}'}) \nonumber \\
 &=&  \frac{1}{2\pi}(\Lambda-p_{F})+\alpha-\alpha(\cos\theta_{\hat{\bmp}}\cos\theta_{\hat{\bmp}'}+\sin\theta_{\hat{\bmp}}\sin\theta_{\hat{\bmp}'}). 
\end{eqnarray}
From Fig. 7, we find $\alpha\sim 0.0284$. 
Then we obtain the expression for $\tilde{\Gamma}(\bmp,\bmp')$ at low temperatures within the RPA:
\begin{eqnarray}
\tilde{\Gamma}(\bmp,\bmp') &\to& \frac{G}{1-G\Pi(0,p_{F},p_{F},\cos\theta_{\hat{\bmp}\cdot\hat{\bmp}'})}.   
\end{eqnarray}
To obtain the BS equation for specific symmetry of pairing, 
we employ the angular decomposion to $\chi(\bmp)$ and $\tilde{\Gamma}(\bmp,\bmp')$. 
The two-dimensional angular momentum eigenfunction is given as
\begin{eqnarray}
y_{l}(\theta) &=& \frac{1}{\sqrt{2\pi}}e^{il\theta},
\end{eqnarray}
where, $l$ is an angular momentum quantum number in a two-dimensional system. 
We only consider the time-reversal invariant pairings because of simplicity. 
Therefore, we take a linear combination:
\begin{eqnarray}
\frac{1}{\sqrt{2}}\frac{1}{\sqrt{2\pi}}(e^{il\theta}+e^{-il\theta}) &=& \frac{1}{\sqrt{\pi}}\cos l\theta.
\end{eqnarray}
Each component of the amplitude is decomposed by this function:
\begin{eqnarray}
\chi^{S}(\bmp) &=& \sum_{l:even}\chi^{S}(|\bmp|)_{l}\frac{1}{\sqrt{\pi}}\cos l\theta_{\hat{\bmp}} \to \sum_{l:even}\chi^{S}_{l}\frac{1}{\sqrt{\pi}}\cos l\theta_{\hat{\bmp}}, \\
\chi^{V}_{0}(\bmp) &=& \sum_{l:odd}\chi^{V}_{0}(|\bmp|)_{l}\frac{1}{\sqrt{\pi}}\cos l\theta_{\hat{\bmp}} \to \sum_{l:odd}(\chi^{V}_{0})_{l}\frac{1}{\sqrt{\pi}}\cos l\theta_{\hat{\bmp}}, \\
|\vec{\chi}^{V}(\bmp)| &=& \sum_{j:odd}|\vec{\chi}^{V}(|\bmp|)|_{j}\frac{1}{\sqrt{\pi}}\cos j\theta_{\hat{\bmp}} \to \sum_{j:odd}|\vec{\chi}^{V}|_{j}\frac{1}{\sqrt{\pi}}\cos j\theta_{\hat{\bmp}},\qquad (j=l\pm1). 
\end{eqnarray}
Here, we negrect $|\bmp|$ dependence of the amplitudes for weak coupling approximation: 
$|\chi(|\bmp|)|_{l}\approx |\chi|_{l}$. Because of (68), we choose even $l$ for scalar in Eq.(79), 
while we choose odd $l$ ($j$) for vector pairings in Eqs.(80) and (81). 
The irreducible vertex part is also decomposed as follows:
\begin{eqnarray}
\frac{1}{G^{-1}-(\Lambda-p_{F})/2\pi-\alpha+\alpha\cos\theta_{\hat{\bmp}\cdot\hat{\bmp}'}} &=& \frac{1}{G^{-1}-(\Lambda-p_{F})/2\pi-\alpha} \nonumber \\
& & -\frac{\alpha}{(G^{-1}-(\Lambda-p_{F})/2\pi-\alpha)^{2}}\cos\theta_{\hat{\bmp}\cdot\hat{\bmp}'} \nonumber \\
& & +\frac{\alpha^{2}}{(G^{-1}-(\Lambda-p_{F})/2\pi-\alpha)^{3}}\cos^{2}\theta_{\hat{\bmp}\cdot\hat{\bmp}'} \nonumber \\
& & -\frac{\alpha^{3}}{(G^{-1}-(\Lambda-p_{F})/2\pi-\alpha)^{4}}\cos^{3}\theta_{\hat{\bmp}\cdot\hat{\bmp}'} \nonumber \\
& & +\cdots. 
\end{eqnarray}
Therefore, when we decide the quantum number $l$ ( or $j$ ) of a BS amplitude, 
the components coupled with the effective interaction $\tilde{\Gamma}(\bmp,\bmp')$ 
are determined {\it a priori} : From the form of the expansion given in Eq. (82), 
we find the fact that, 1st, 3rd,... terms in Eq. (82) couple only with components of even $l$, 
while 2nd, 4th,... terms in Eq. (82) couple with components of odd $l(j)$. 
It is an important fact that, the sign in each term in the expansion for $\tilde{\Gamma}$ 
alternates between plus and minus.

Based on the preparation given above, 
we obtain the BS equations for specific $l$ ( or $j$ ) of several types of pairings. 
For the case of the scalar:
\begin{eqnarray}
1 &=& \frac{1}{8\pi^{2}}\tilde{I} \int^{\Lambda}_{0}p'dp' \nonumber \\ 
& & \times\Bigl(\frac{1}{2(|\bmp'|-\mu)}\tanh\frac{\beta}{2}(|\bmp'|-\mu)+\frac{1}{2(|\bmp'|+\mu)}\tanh\frac{\beta}{2}(|\bmp'|+\mu)\Bigr).
\end{eqnarray}
For the 0th-component of vector:
\begin{eqnarray}
1 &=& \frac{1}{8\pi^{2}}\tilde{I} \int^{\Lambda}_{0}p'dp' \frac{1}{2\mu}\Bigl(-\tanh\frac{\beta}{2}(|\bmp'|-\mu)+\tanh\frac{\beta}{2}(|\bmp'|+\mu)\Bigr).
\end{eqnarray}
Here $\tilde{I}$ is determined as
\begin{eqnarray}
\tilde{I} &=& \int^{2\pi}_{0}d\theta_{1}\int^{2\pi}_{0}d\theta_{2}\frac{\cos l\theta_{1}\cos l\theta_{2}}{G^{-1}-(\Lambda-p_{F})/2\pi-\alpha+\alpha\cos\theta_{12}},
\end{eqnarray}
where $\theta_{1}=\theta_{\hat{\bmp}}$, $\theta_{2}=\theta_{\hat{\bmp}'}$ 
and $\theta_{12}=\theta_{1}-\theta_{2}$. 
In principle, we have to choose $\Lambda$ in our BS equations (83) and (84) as 
the same value for the polarization in Eq.(74). By using the expansion (82), 
when $G<0$, we find the following fact: In the angular integration of $\tilde{I}$, 
there are only negative contribution for $\tilde{I}$ ( $\tilde{I}<0$ ) of the scalar case, 
while there are only positive contribution for $\tilde{I}$ ( $\tilde{I}>0$ ) of the vector case. 
Therefore, from Eq.(83), we recognize that there is no solution in the scalar pairing. 
From the same reason given in the discussion for Eq. (62), 
we cannot find solution for the 0th-component of vector case. For the spatial-component of vector, we get
\begin{eqnarray}
-1 &=& \frac{1}{8\pi^{2}}\int^{\Lambda}_{0} p'dp'\int^{2\pi}_{0} d\theta_{1} \int^{2\pi}_{0} d\theta_{2}  \nonumber \\
& & \times \frac{\cos j\theta_{1}\cos j\theta_{2}}{G^{-1}-(\Lambda-p_{F})/2\pi-\alpha+\alpha\cos\theta_{12}} \nonumber \\
& & \times \frac{1}{2\mu|\bmp'|}\Bigl(\frac{|\bmp'|^{2}\cos^{2}\theta_{2}-\mu|\bmp'|}{|\bmp'|-\mu}\tanh\frac{\beta}{2}(|\bmp'|-\mu)-\frac{|\bmp'|^{2}\cos^{2}\theta_{2}+\mu|\bmp'|}{|\bmp'|+\mu}\tanh\frac{\beta}{2}(|\bmp'|+\mu)\Bigr). 
\end{eqnarray}
In Eq. (86), the angular integration gives the negative quantity, 
but the integrand of momentum integration is also always negative with respect to the variation of $T$. 
Therefore, it is impossible to find $T_{c}$, and we have no solution in the spatial-component of vector. 
The summary of the results of the solutions of various types of our gap equations and BS equations 
are presented in table I. Only the scalar pairing in the attractive interaction can have nontrivial 
solution. In our treatment ( BS-RPA ), we could not find a KL effect in our model. 
Baranov et al.~[60] obtained a conclusion that no pairing instability arises in a two-space-dimensional 
nonrelativistic model with second order perturbation. 
Chubukov performed a calculation beyond second order: 
He included a vertex correction, and showed that the KL effect arises in 
the two-space-dimensional Fermi liquid~[47]. 
Our negative result of the KL effect is based on both the single conical dispersion model 
and the angular dependence of the irredicible vertex part estimated by RPA. 
Therefore, our result does not deny the presence of the KL effect in (2+1)-dimensional systems completely. 
In fact, Gonz\'alez et al. showed the presence of the KL effect in graphite, 
by taking into account the interaction between two inequivalent Fermi points carefully~[32]. 
On the other hand, as discussed in the introduction, 
we have set the purpose of this paper just to examine the lower-band effect in superconductivity. 
For the purpose, we use a single relativistic-dispersion model with $\mu\ne0$. 
The situation of our model is different from that of the work of Gonz\'alez et al.

\section{Concluding Remarks}

In this paper, we have investigated the relativistic model of 
two-band superconductivity in (2+1)-dimension. 
After we have introduced the model Lagrangian, 
we have derived the (2+1)-dimensional relativistic Gor'kov equation for 
the pairing problem under the four-fermion contact attractive interaction. 
We have performed the group-theoretical consideration of the gap functions. 
The characteristic aspects of the gap functions were revealed. 
By using the Gor'kov formalism, we have derived the gap equations for several types of pairings, 
and have solved them numerically. 
We have understood quantitatively the effect of the lower band in the two-band superconductivity: 
We have found the lower band enhances the superconductivity. 
We also have examined the pairing problem under the repulsive interaction ( the KL effect ), 
by using the BS formalism.

Now, we discuss some remaining problems and/or further possible investigations in our theory.
It is interesting to perform the calculations of the response function or polarization function 
under the presence of electromagnetic field, because of the reason that 
those functions can give the Chern-Simons ( CS ) term or not. 
Goryo and Ishikawa discussed the induction of the CS term in (2+1)-dimensional nonrelativistic theory, 
with parity and time-reversal violating superconductors~[61]. 
Hosotani showed the (2+1)-dimensional QED with the CS term dynamically 
generate a magenetic field~[62]. 
Miransky et al. studied the fact that an external magnetic field enhances 
a fermion dynamical mass, and this phenomenon is universal in 
any models of (2+1) and (3+1) dimensional field theories~[63]. 
We suppose these studies should have an intrinsic relation. 
We have a plan to investigate these physics in our model in near future.

\acknowledgments

The author would like to express his gratitude sincerely to Professor Yoichiro Nambu, 
for his enlightening discussions and comments. 
Thanks are also due to Professor Hiroshi Toki, for his critical reading of this paper.
The author is also grateful to Professors H. Akai, T. Morinari, H. Nagao, M. Nakano, 
H. Tsunetsugu, K. Yamaguchi, D. Yamaki and S. Yamanaka, for their encouragements during this work.

\begin{figure}

\caption{(a) The structure of the honeycomb lattice. A and B denote the two different sublattice sites. (b) A schematic figure of the band structure of the honeycomb lattice. By symmetry, the hexagonal two-dimensional Brillouin zone of it has two degeneracy points. In the case of graphite, the Fermi levels locate on these points. In this paper, we use a Dirac fermion model with $\mu\ne 0$. Our theory can be applied to electron-doped graphite, etc.}

\vspace{10mm}

\caption{A schematic figure of the relativistic dispersion in a two-dimensional system. $\epsilon_{F}$ denotes the Fermi energy of the system.}

\vspace{10mm}

\caption{The $G$ dependence of the pairing gap at $T=0$. We set $\Lambda=2k_{F}=2\mu$ and $\mu=1$.}

\vspace{10mm}

\caption{The temperature dependence of the pairing gap. We set $G=2\mu^{-1}$, $\mu=1$ and $\Lambda/2\mu=1$.}

\vspace{10mm}

\caption{(a) The diagrammatic representation for the two-particle Green's function $G^{(2)}$. $\tilde{\Gamma}$ denotes the irreducible vertex part. (b) The diagram for the irreducible vertex part within the RPA. The solid points represent the bare vertices $G$.}

\vspace{10mm}

\caption{The $q$-dependence of the static polarization $\Pi(0,q)$ under various temperature. Here we set $\mu=1$ and $\Lambda/2\mu=1$.}

\vspace{10mm}

\caption{The $\theta$-dependence of the static polarization $\Pi(0,k_{F},k_{F},\cos\theta)$ under various temperature. Here we set $\mu=1$ and $\Lambda/2\mu=1$.}

\end{figure}

\begin{table}
\caption{List of solutions for various pairings.}
\begin{tabular}{ccc}
pairing symmetry & Gor'kov ( $G>0$ ) & Bethe-Salpeter-RPA ( $G<0$ ) \\
\hline
scalar & possible & no solution \\
0th-component of vector & no solution & no solution \\
spatial-component of vector & no solution & no solution \\
\end{tabular}
\end{table}


\begin{thebibliography}{99}
\bibitem{kagome} 
The kagome lattice is a three-band system, but it has conical dispersions similar to the case of the honeycomb lattice.
\end{thebibliography}
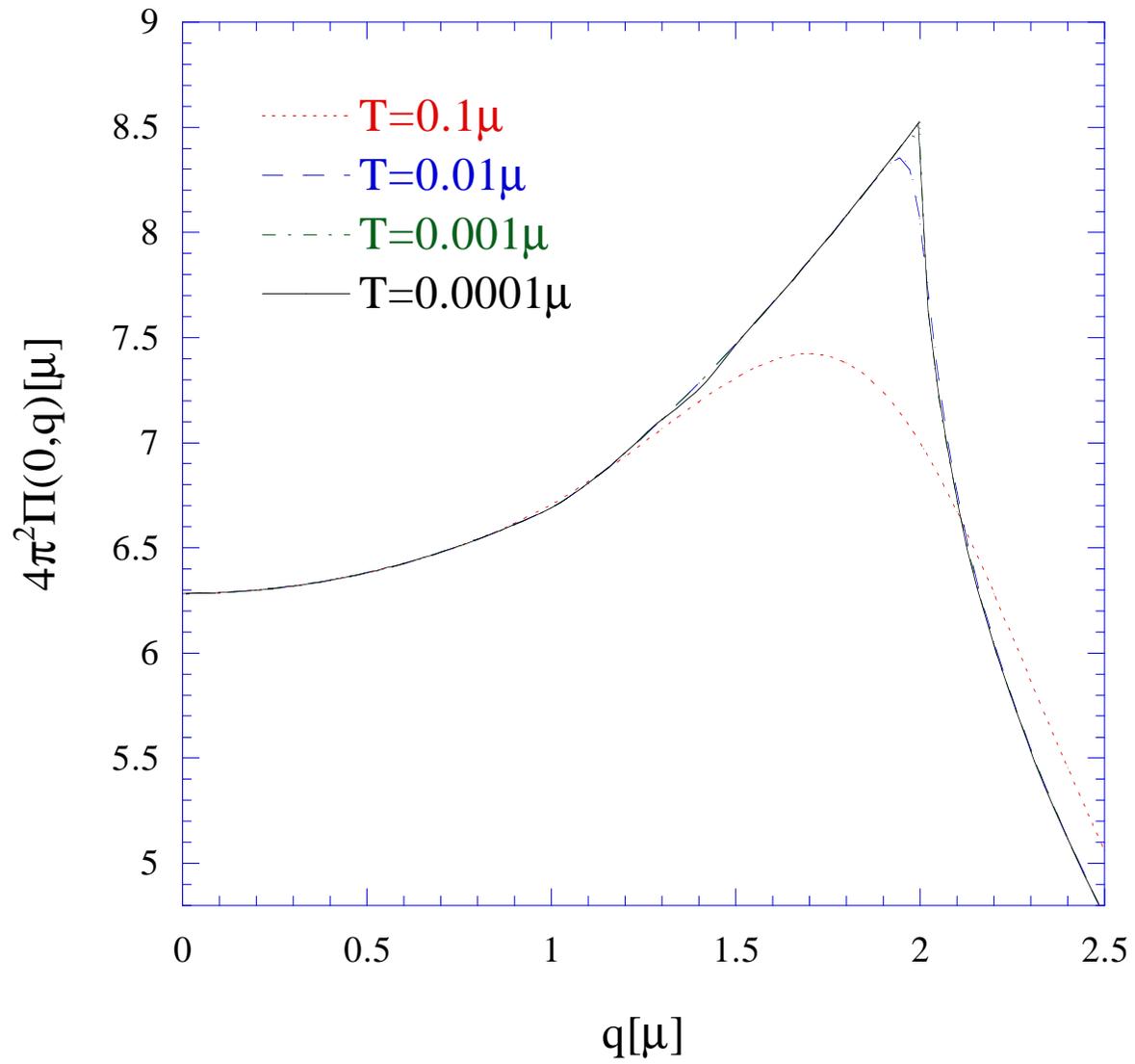
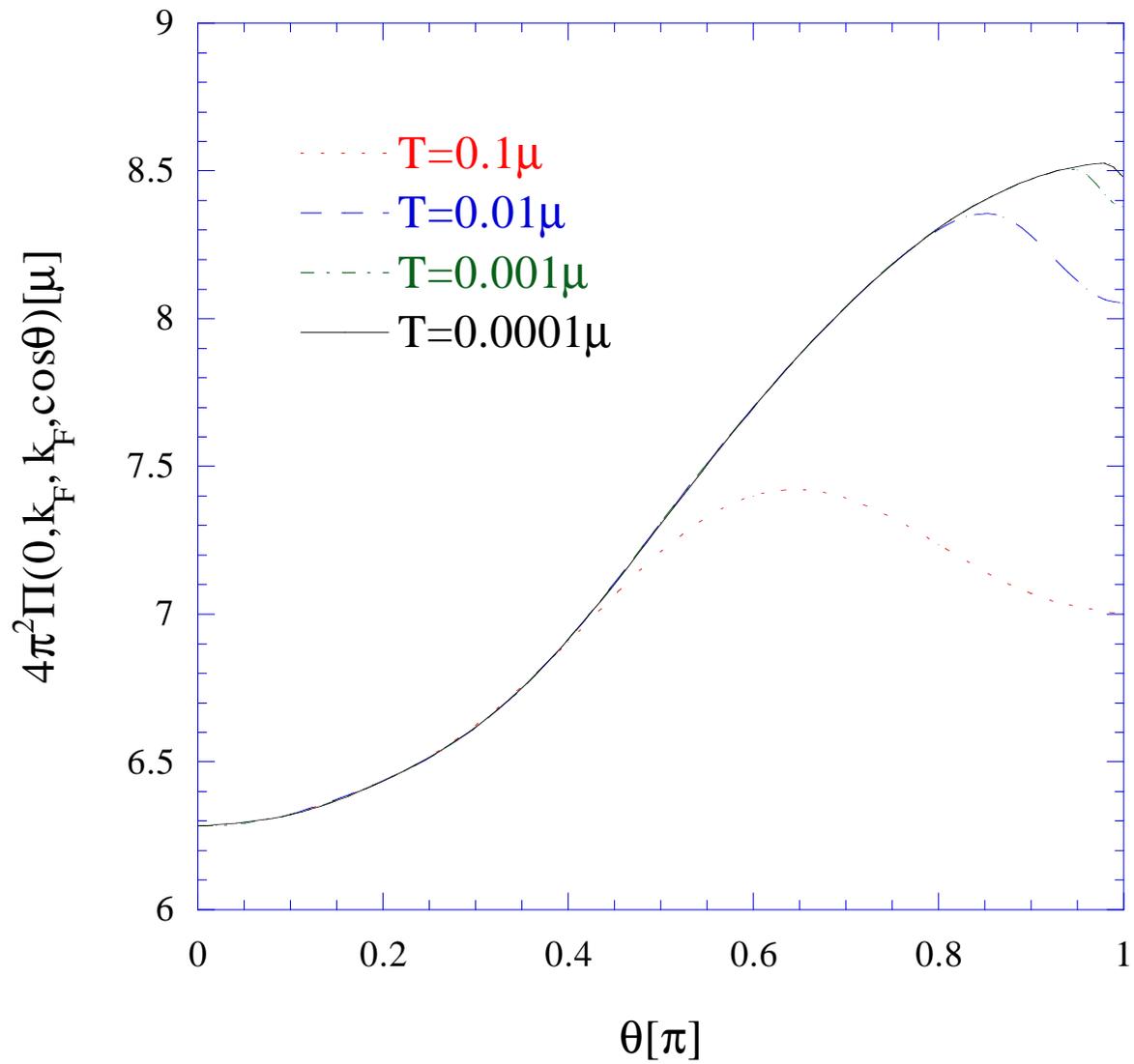
\end{document}